
\nopagenumbers

\def\dl{{[\hskip -1.5pt [}}         
\def\dr{{]\hskip -1.5pt ]}}         

\input amstex
\loadbold
\loadeusm
\input epsf

\documentstyle{amsppt}
\TagsOnRight

\hoffset=.15truein
\hsize=6.2truein
\voffset=-.25truein         
\vsize=9.0truein
\parindent=20pt

\overfullrule=0pt

\topmatter
\title The warp drive and antigravity \endtitle
\author Homer G. Ellis \endauthor

\address
\vskip -30pt
\rm
$$
\alignat 2
\text{First version:  }& @,\text{February, 1996}& \\
\vspace{-3pt}
\text{Revised:  }& @,@,@,@,@,@,\text{\phantom{Febr}July, 1996}& \\
\vspace{-3pt}
\text{Revised:  }& @,@,@,@,\text{\phantom{Fe}March, 2001}&
\endalignat
$$
\vskip -5pt
\indent Homer G. Ellis\newline
\indent Department of Mathematics\newline
\indent University of Colorado at Boulder\newline
\indent 395 UCB\newline
\indent Boulder, Colorado  80309-0395\newline\newline
\indent Telephone: (303) 492-7754 (office); (303) 499-4027 (home)\newline
\indent Email:  ellis\@euclid.colorado.edu\newline
\indent Fax: (303) 492-7707
\endaddress

\abstract
\it The warp drive envisioned by Alcubierre that can move a spaceship faster
than light can, with modification, levitate it as if it were lighter than
light, even allow it to go below a black hole's horizon and return unscathed.
Wormhole-like versions of the author's `drainhole' (1973) {\rm might}
provide the drive, in the form of a by-pass of the spaceship composed of a
multitude of tiny topological tunnels.  The by-pass would divert the
gravitational `ether' into a sink covering part of the spaceship's hull,
connected by the tunnels to a source covering the remainder of the hull, to
produce an ether flow like that of a river that disappears underground only
to spring forth at a point downstream.  This diversion would effectively
shield the spaceship from external gravity.
\endabstract

\endtopmatter

\document

\vskip -15pt

In a letter that appeared in 1994 in the journal {\it Classical and Quantum
Gravity} Alcubierre exhibited a space-time metric that describes a surprising
phenomenon occurring in a flat, euclidean space:  a spherical region
of the space glides along geodesically with a prescribed velocity
$\bold v_{\roman s} (t)$ as if it were a (practically) rigid body unattached
to the remainder of space [1].  The velocity, directed along the $x$ axis,
is arbitrary as to magnitude and time dependence.  In particular, the speed
of the moving region can be anything from zero to many times the speed of a
light pulse traveling on a parallel track outside the sphere.  The ability
so to select $\bold v_{\roman s} (t)$ makes possible long trips in short
times at high speeds.  The times are measured to be the same by travelers
inside the sphere and observers outside the sphere.  The speeds are those
measured by the external observers.  The travelers, unless they look at
fixed points outside the sphere, will be unaware that they are moving, for
everything inside, light included, behaves as if the sphere were at rest.

This somewhat counterintuitive motion of the spherical region involves a
distortion of space-time highly localized at the region's boundary.  As
Alcubierre noted, a mechanism for producing that distortion, however it
might be designed, would fit well the picturesque name `warp drive'
familiar from science fiction.  In this paper I shall show that such a
warp drive can be made to serve as an antigravity device, and shall describe
a topological design that causes the idea of constructing one to seem a
little less far-fetched than conventional wisdom would suggest.

The space-time metric that Alcubierre exhibited achieves its effect by
replacing the zero velocity of the motionless points of empty space by the
translational velocity $\bold v_{\roman s} (t)$, but only (to a near
approximation) inside a sphere of radius $R$, which sphere we may for
purposes of the present discussion take to be the skin of a spaceship
propelled by the warp drive (with us in it, let us say).  This sphere
centers on the point at $\bold x_{\roman s} (t)$, which moves so that
$d\,[\bold x_{\roman s} (t)]/dt = \bold v_{\roman s} (t)$ at all times.  The
restriction of the motion to the $x$ direction, introduced for simplicity's
sake, may be dispensed with.  The space-time produced by this distortion of
flat Minkowski space-time has then the proper-time line element
$d\tau$ given by
$$
d\tau^2 = dt^2 - | d\bold x - \bold u (t, \bold x) \, dt |^2,
\tag1
$$
where
$\bold u (t, \bold x) := \bold v_{\roman s} (t) f(r_{\roman s}(t, \bold x))$
and
$r_{\roman s} (t, \bold x) := | \bold x - \bold x_{\roman s} (t) |$, the
function $f$ being defined by
$$
f(r) := \displaystyle {\tanh (\sigma (r + R)) - \tanh (\sigma (r - R))
                            \over 2 \tanh (\sigma R)},
\tag2
$$
so that, as $\sigma \to \infty$, $f(r)$ tends to 1 if $| r | < R$ but to 0
if $| r | > R$.  Every space-time path with 4-velocity
$\dl\, 1, \bold u \,\dr$, thus with 3-velocity $d\bold x/dt = \bold u$,
is geodesic; outside the spaceship $\bold u \approx \bold 0$, inside
$\bold u \approx \bold v_{\roman s}$ ($= \bold v_{\roman s}$ at the center).
The function $f$ interpolates between the exterior velocity $\bold 0$ and the
interior velocity $\bold v_{\roman s}$, abruptly replacing the one with the
other at the spaceship's skinny boundary, where
$r_{\roman s} (t, \bold x) = R$.

If the points of space are themselves not sitting still, rather are
streaming along with flow velocity $\bold v (t, \bold x)$, the same velocity
interpolation takes the form
$$
\bold u (t, \bold x) :=
              \bold v_{\roman a} (t, \bold x) [1 - f(r_{\roman s}(t, \bold x))]
                   + \bold v_{\roman s} (t) f(r_{\roman s}(t, \bold x)),
\tag3
$$
with $\bold v_{\roman a}$, the ambient velocity, equal to $\bold v$.  The
velocity field $\dl\, 1, \bold u \,\dr$ remains geodesic, but now outside
the spaceship $\bold u \approx \bold v_{\roman a}$, while inside still
$\bold u \approx \bold v_{\roman s}$.  This would be the situation if the
spaceship were immersed in a gravitational field representable by a metric
$$
dt^2 - | d\bold x - \bold v (t, \bold x) \, dt |^2,
\tag4
$$
because for this metric the geodesic 4-velocity
$\dl\, 1, \bold v \,\dr$ can be interpreted as that of a point of
space moving with 3-velocity $\bold v$ (with respect to an immobile
background space, one has to say).  Using our warp drive to distort this
metric to that of equation (1), we can, with the choice of
$\bold v_{\roman s}$ at our disposal, navigate freely in the gravitational
field, even stop at will to inspect our environs.  The Schwarzschild field
has such a representation, for the Schwarzschild metric of an object of
active gravitational mass $m$, namely
$$
(1 - 2 m/\rho) \, dT^2 - (1 - 2 m/\rho)^{-1} d\rho^2
                         - \rho^2 d\vartheta^2
                         - \rho^2 (\sin \vartheta)^2 d\varphi^2,
\tag5
$$
is brought by the transformation
$T = t - \int \! \sqrt{2 m/\rho \,} \, (1 - 2 m/\rho)^{-1} \, d\rho$
to the form
$$
dt^2 - (d\rho + \sqrt{2 m/\rho \,} \, dt)^2
       - \rho^2 d\vartheta^2
       - \rho^2 (\sin \vartheta)^2 d\varphi^2,
\tag6
$$
and then, upon conversion of the spherical coordinates
$\dl\, \rho, \vartheta, \varphi \,\dr$ to cartesian, to the form (4)
with $\bold v (t, \bold x) = \bold v_{\roman {Sch}} (t, \bold x) :=
-\sqrt{2 m/|\bold x|} \, (\bold x/|\bold x|)$.  In this representation the
acceleration of a radially moving test particle is
$-\nabla (-{1\over2} |\bold v_{\roman {Sch}}|^2)$ ($= -m/|\bold x|^2$),
so $-{1\over2} |\bold v_{\roman {Sch}}|^2$ plays the role of Newtonian
gravitational potential.

With the ambient gravitational field thus canceled inside the spaceship
we normally will float about, bouncing off the bulkheads.  When the novelty
of this wears off, we can gain the illusion of {\it terra firma} under our
feet by simulating the presence of Earth beneath the spaceship.  All that
is necessary is to modify the interior velocity to
$\bold v_{\roman s} (t) + \bold v_{\! g} (t, \bold x)$, with $\bold v_{\! g}$
defined by
$$
\bold v_{\! g} (t, \bold x) :=
  - \sqrt{2 g (R - (\bold x - \bold x_{\roman s} (t))
                              \bold \cdot \bold n)} \,\, \bold n,
\tag7
$$
in which $\bold n$, the `upward pointing' unit vector normal to the
spaceship's deck, is presumed to have a fixed direction.  That will give
everything inside the spaceship and `above' the deck an acceleration
$-g \bold n$ toward the deck.  This acceleration, attributable to the spatial
non-uniformity of $\bold v_{\! g} (t, \bold x)$, will be with respect to the
internal space of the spaceship, unlike $d\,[\bold v_{\roman s} (t)]/dt$,
which, being spatially uniform inside and not in effect outside, is an
acceleration with respect only to the space outside the ship and therefore
goes unnoticed within.

If the Schwarzschild object is a black hole, we can with our warp drive go
below its horizon at $\rho = 2 m$ and come back up unscathed.  The drive
has only to work a little harder just below the horizon, where
$| \bold v_{\roman a} | = | \bold v_{\roman {Sch}} | > 1$, than just above,
where $| \bold v_{\roman a} | = | \bold v_{\roman {Sch}} | < 1$.  To escape
from under the horizon we need only command the drive to give
$\bold v_{\roman s} (t)$ a non-zero outward component.  Nor will it matter
that the black hole is rotating.  From its expression in Boyer-Lindquist
`Schwarzschild-like' coordinates
$\dl\, T, \rho, \vartheta, \varphi \,\dr$ [2] the Kerr rotating black
hole metric transforms to
$$
\align
dt^2
 &- \left[ 1 - a^2 (\sin \vartheta)^2 \frac{\Sigma \Delta - (2 m \rho)^2}
                                           {\Sigma \Delta^2} \right]
     \left( d\rho + \frac{\sqrt{2 m \rho \, (\rho^2 + a^2)}}
                           {\Sigma} \, dt \right)^2
  - \Sigma \, d\vartheta^2 \\
 &- \left[ \frac{4 a m \rho \, (\sin \vartheta)^2
            \sqrt{2 m \rho \, (\rho^2 + a^2)}} {\Sigma \Delta} \right]
    \left( d\rho + \frac{\sqrt{2 m \rho \, (\rho^2 + a^2)}}
                          {\Sigma} \, dt \right)
    \left( d\varphi - \frac{2 a m \rho}
                          {\Sigma \Delta} \, dt \right) \\
 &- \left[ (\rho^2 + a^2) (\sin \vartheta)^2
             + \frac{2 m \rho} {\Sigma} \, a^2 (\sin \vartheta)^4 \right]
    \left( d\varphi - \frac{2 a m \rho} {\Sigma \Delta} \, dt \right)^2,
\tag8
\endalign
$$
with $\Sigma = \rho^2 + a^2 (\cos \vartheta)^2$ and
$\Delta = \rho^2 - 2 m \rho + a^2$, when the substitution
$T = t - \int \! \sqrt{2 m \rho \, (\rho^2 + a^2)} \,\, 
\Delta^{-1} \,\, d\rho$ is made [3].  This has the form
$$
dt^2 - \gamma_{i j} (x) (dx^i - v^i (x) \, dt)
                          (dx^j - v^j (x) \, dt),
\tag9
$$
with $\dl\, x^i \,\dr = \dl\, \rho, \vartheta, \varphi \,\dr$,
$v^\rho = -\sqrt{2 m \rho \, (\rho^2 + a^2)} \big/ \Sigma$,
$v^\vartheta = 0$, and $v^\varphi = {2 a m \rho} \big/ {\Sigma \Delta}$.
Although the spatial geometry described by the metric $\gamma_{i j}$ is not
flat if $a \neq 0$, velocity interpolation like that above will replace the
velocity $v$ by a velocity $v_{\roman s}$ inside the spherical spaceship,
the result being a metric
$$
dt^2 - \gamma_{i j} (x) (dx^i - u^i (t, x) \, dt)
                          (dx^j - u^j (t, x) \, dt),
\tag10
$$
where
$$
u^i (t, x) := v_{\roman a}\!{}^i (x) [1 - f(r_{\roman s} (t, x))]
                 + v_{\roman s}\!{}^i (t) f(r_{\roman s} (t, x)),
\tag11
$$
$v_{\roman a} = v$, and $r_{\roman s} (t, x)$ is the geodesic distance from
the center of the sphere at $x_{\roman s} (t)$ to the point at $x$,
measured by the metric $\gamma_{i j}$.  As in the euclidean case, the
velocity field $\partial_t + u^i \partial_i$ is geodesic, so the points of
space inside the ship will move in concert almost as a rigid body with
velocity $v_{\roman s}$.

For the non-rotating black hole the volume expansion $\theta$ at time $t$
of the geodesic velocity field $\partial_t + u^i \partial_i$ is just the 
flat-space divergence of $\bold u$, calculated from equation (3) with
$\bold v_{\roman a} = \bold v_{\roman {Sch}}$.  Figure 1 is a shaded contour
plot of $\theta$ restricted to an arbitrary plane through the $x$ axis,
with $\sigma = 8$, $R = 1$, $m = 5$,
$\bold x_{\roman s} (t) = \dl\, 4, 0, 0 \,\dr$, and
$\bold v_{\roman s} (t) = \bold 0$.  Lowest (negative) values of $\theta$
show as black, highest (positive) values as white.  The spaceship is holding
its position with $|\bold x| = 4$, well inside the black hole's event horizon
at $|\bold x| = 2 m = 10$.  Figure 2 is like Figure 1, but with the plane
fixed as the $xy$ plane and $\bold v_{\roman s} (t) = \dl\, 0, 2, 0 \,\dr$.
The spaceship is at `periholion', passing the black hole below the horizon
along a path (vertical in the figure) that is tangential to the sphere of
symmetry at $|\bold x| = 4$. These plots are analogs of the surface plot of
$\theta$ shown in Figure 1 of Alcubierre's letter.

\epsfxsize=6.9truein
\epsffile[100 460 600 701]{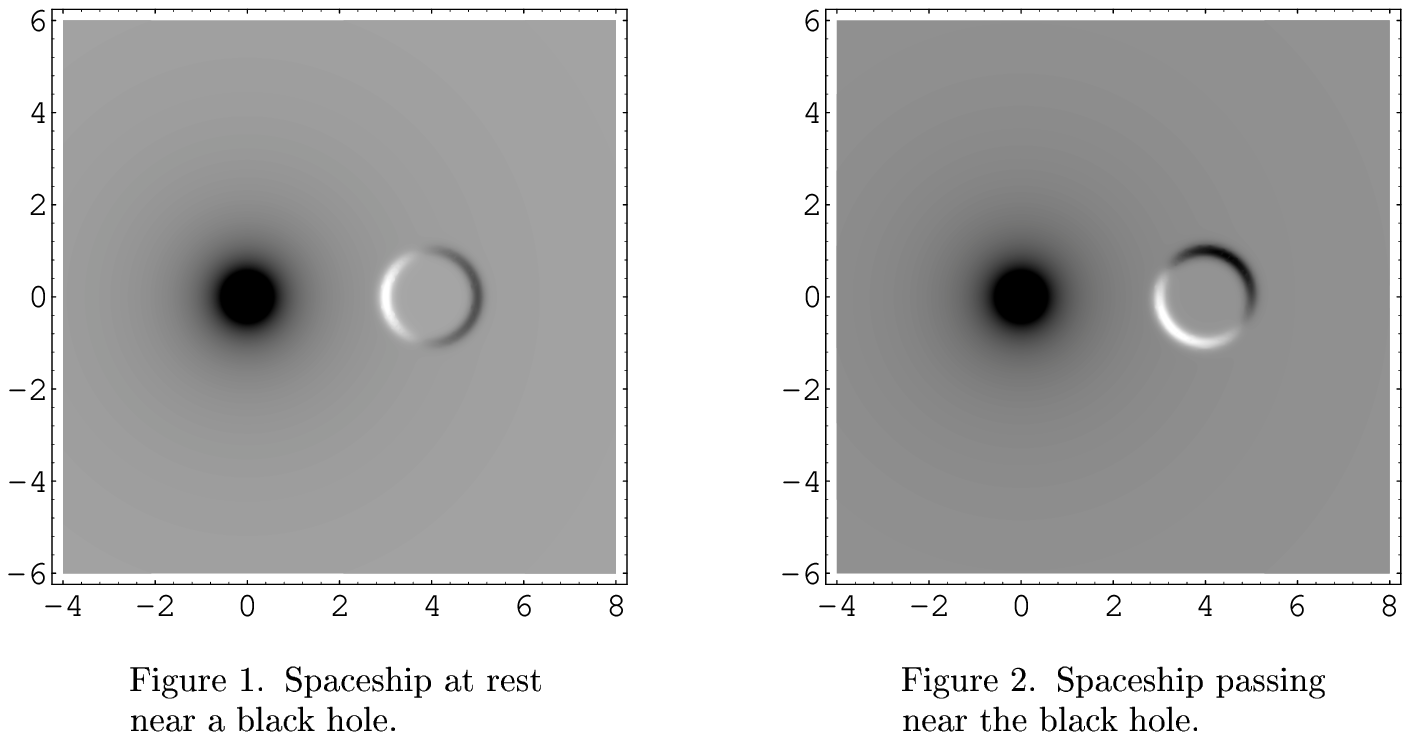}

\vskip -5pt

Alcubierre reasoned that his spaceship moves along with velocity
$\bold v_{\roman s}$ because some unspecified mechanism involving `exotic'
matter continuously shrinks space in front of it (where $\theta < 0$) and
expands space behind it (where $\theta > 0$).  That interpretation extended
to the current development would say that the black and the white regions
of the contour plots are places where space is shrinking (black) or expanding
(white).  An alternative way of describing what is happening arises from an
idea in [4].  In that paper I attributed the gravity of the Schwarzschild
field to the ``internal, relative motions'' of a ``more or less substantial
`ether', pervading all of space-time''.  The velocity $\bold v_{\roman e}$
of this `ether' is just the velocity $\bold v_{\roman {Sch}}$ given above,
which one recognizes as the velocity of an observer falling freely from rest
at $\bold x = \bold \infty$.  The role of gravitational potential thus falls
to the scalar field $-{1\over2} |\bold v_{\roman e}|^2$, the negative of the
`specific kinetic energy' of the ether.

Application of this `ether flow' picture to the present discussion would
suggest that the black regions in Figure 1 indicate the presence of ether
sinks, one at the Schwarzschild singularity and one spread over the upper
and some of the lower hemisphere of the spaceship's skin, whereas the white
region indicates an ether source spread over the remainder of the lower
hemisphere of the skin.  The enveloping of the spaceship in this way by an
ether sink and an ether source of just the right strengths allows the ether
inside the spaceship to remain at rest, aloof from the headlong rush of the
surrounding ether into the sink at $\bold x = \bold 0$.  Modification of the
strengths and locations of the enveloping sink and source as in Figure 2
produces an outward velocity of the sink, the source, and the ether inside
the spaceship, thus permits the ship to escape the black hole.  Far away
from gravitating matter, as in Alcubierre's example,
$\bold v_{\roman e} \approx \bold 0$.  The ether outside the spaceship is
nearly at rest, but that ahead is drawn continuously into a sink region at the
front of the spaceship at a rate determined by the forward speed of the ship,
while that behind is added to at a similar rate out of a source region in
back.  The ether inside the spaceship moves along in step with the sink and
the source.  Having no internal, relative motion, it produces no gravity.

Just what the `ether' might be is a question not addressed in [4].  Whether
one thinks of the ether as a fluid of some kind spread throughout space, or
as space itself flowing in time, or as just a convenient fiction is largely
a matter of taste --- the mathematics is the same in every case.  The idea
of {\it something} flowing does, however, suggest a way for the warp drive
to be brought into existence.  Creation of a $\theta < 0$ and a
$\theta > 0$ region independent of one another, enveloping the spaceship
and able to drive it, would require accumulation of a considerable amount
of attractive active gravitational mass in front of the ship and a
hard-to-imagine independent accumulation of a similar amount of repulsive
active gravitational mass behind it.  But should these regions be an ether
source and an ether sink, connected so that the ether disappearing into the
sink reappears at the source in the manner that a river disappears
underground only to spring forth at a point downstream, then perhaps they
can be created more easily.

In the ether flow paper, to remedy the undesirable destruction of ether by
the singularity at the black hole sink, I proposed an alternative to the
Schwarzschild black hole, and termed it a `drainhole'.  The drainhole is a
static solution of the usual coupled Einstein--scalar-field equations, but
with nonstandard coupling polarity.  Its metric, dependent on a mass
parameter $m$ and a parameter $n > |m|$, takes the spherically symmetric,
radial ether flow form, analogous to the form (6) of the Schwarzschild
metric in spherical coordinates,
$$
dt^2 - (d\rho - v_{\roman e}\!{}^\rho (\rho) dt)^2
    - r^2(\rho) d\vartheta^2 - r^2(\rho) (\sin \vartheta)^2 d\varphi^2,
\tag12
$$
where
$$
r (\rho) =
 \sqrt{\rho^2 - 2 m \rho + n^2} \exp \left({m \over n} \alpha (\rho)\right)
\tag13
$$
and
$$
v_{\roman e}\!{}^\rho (\rho) = -\text{sgn}(m)\left[1 - \exp\left(-{2m \over n}
                                         \alpha (\rho)\right)\right]^{1/2},
\tag14
$$
with
$$
\alpha (\rho) = {n \over \sqrt{n^2 - m^2}} \left[{\pi \over 2} - \tan^{-1}
                      \left(\frac{\rho - m}{\sqrt{n^2 - m^2}}\right)\right]
\tag15
$$
\vskip 5pt
\noindent ($\rho$ here corresponds to $\rho + m$ in [4], and ranges from
$-\infty$ to $\infty$).  If $m > 0$, then $r(\rho) \sim \rho$ and
$v_{\roman e}\!{}^\rho (\rho) \sim -\sqrt{2 m/\rho}$, as $\rho \to \infty$, so
the drainhole's behavior is asymptotic to that of the black hole as
$\rho \to \infty$.  Unlike the black hole, however, the drainhole is
geodesically complete.  Where the black hole has two asymptotically flat
outer regions, connected for a short time by a `throat' at the horizon, and
two inner regions, each with a central singularity where curvatures become
infinite, the drainhole has only two asymptotically flat regions (one where
$\rho \to \infty$, the other where $\rho \to -\infty$) connected by a throat
for as long as the drainhole exists, and has no singularity or horizon at all.

In the black hole the ether flow acceleration is everywhere inward, so the
black hole is gravitationally attractive on both sides.  Contrarily, the
Schwarzschild white hole, whose metric is given by (5) with $m < 0$, can
only repel gravitationally.  It consequently is not susceptible to an ether
flow description, there being no possibility of an observer's falling freely
{\it from} (or {\it to}) {\it rest} at $|\bold x| = \infty$, and therefore no
vector field corresponding to the $\bold v_{\roman {Sch}}$ of the black hole.
Nevertheless, accumulation of small black holes in front of the spaceship
and small white holes, if such could be produced, behind it would create a
warp drive consisting of multiple sinks in front, taking in ether with no
place to put it, and unrelated multiple repellers in back, somehow
fabricating new space out of nothing and pushing it away by a mechanism
unknown.

A drainhole can be thought of as a happy union of a black hole and a white
hole in which the (w)hole is better than the sum of its parts.  The
acceleration of the radially flowing ether in a drainhole is given by
$d^2\! \rho/dt^2 = d\,[v_{\roman e}\!{}^\rho (\rho)]/dt = -m/r^2 (\rho)$.
The drainholes with $m < 0$ are metrically indistinguishable from those
with $m > 0$.  In each the ether flows from one asymptotically flat region,
through the throat, and out into the other, accelerating at every point in
the direction of the flow, which is inward (toward the throat) on the
`high' side, where $\rho/{2 m} > 1$, and outward (away from the throat) on
the `low' side, where $\rho/{2 m} < 1$.  The ether comes out faster than it
went in, and flows the faster the farther out it travels.  Thus the
drainhole appears on the high side as a gravitationally attracting ether
sink, and on the low side as a gravitationally repelling ether source.  Not
only that, but the strength of the repulsion exceeds that of the
attraction, by a percentage calculable as approximately
$\pi m/n$ if $0 < m \ll n$.  What is more, to keep the drainhole throat
open does not require that the ether flow rapidly, or even at all --- the
throat's smallest possible constriction, which occurs when the ether is not
flowing, that is, when $m = 0$ so that $v_{\roman e}\!{}^\rho = 0$, is a
two-sphere of area $4 \pi n^2$.  All of this is established in [4].

Drainholes allowed to evolve can appear and disappear.  In [5] I derived a
solution of the coupled Einstein--scalar-field equations in the form of a
space-time manifold $\eusm M_a$ comprising, if the parameter $a \neq 0$,
two asymptotically flat regions connected by a throat that constricts to a
point and immediately reopens and begins to enlarge (a phenomenon
prefigurative of the `scalar field collapse' studied later in [6--10]).  The
metric is $dt^2 - d\rho^2 - r^2(t, \rho) d\vartheta^2 -
r^2(t, \rho) (\sin \vartheta)^2 d\varphi^2$, with
$r^2(t, \rho) = a^2 t^2 + (1 + a^2) \rho^2$.  Combining the $t > \rho$
region of $\eusm M_a$ for $a \neq 0$ and the $t < \rho$ region of
$\eusm M_0$ produces a drainhole in which the throat is absent when
$t \leq 0$, but present when $t > 0$.  Combining the $t < \rho$ region
of $\eusm M_a$ and the $t > \rho$ region of $\eusm M_0$ produces a
drainhole in which the throat is present when $t < 0$, but absent when
$t \geq 0$.  The ether is at rest, but the existence of analogous
solutions with the ether flowing is plausible.

If small wormhole-like versions of these drainholes could be manufactured
in great numbers, with their high sides distributed over one face of a
closed vessel of spherical (or perhaps spheroidal!) \! shape, and their low
sides spread over the opposite face, the result would be an ether by-pass
of the vessel consisting of a multitude of tiny topological tunnels.  With
such an ether by-pass the vessel could be shielded from the external
gravity embodied in the flow of the ether, and could become a gravity
defying spaceship --- in short, a warp drive would exist.  It is easy
enough to imagine the mathematical existence of such a diverted-ether-flow
space-time configuration (imagine the spatial topology with the velocity
field of the flow painted on), but concrete physical existence is another
matter.  One is tempted to speculate that something as conceptually simple
as a generator of particle-antiparticle pairs, coupled to an accelerator to
separate the particles and the antiparticles and spread them over opposite
faces of the vessel, would do the job.  This, though, presumes more than is
known by Earthlings about the active gravitational masses of such things as
electrons and positrons.

Whatever the means by which such a warp drive might be realized --- if
realized one ever should be --- it is worthy of note that once in existence
the drive would be able to levitate a heavily loaded vessel almost as
easily as it would the vessel alone.  To levitate an empty vessel on
Earth, whose active gravitational mass $M_{\roman E}$ is treated as
concentrated at its center a distance $R_{\roman E}$ below its surface, the
drive must merely provide a by-pass for ether flowing downward with velocity
$\sqrt{2 M_{\roman E}/R_{\roman E}}$, which is Earth's escape velocity of
about $1.1 \times 10^4$ m s$^{-1}$, and acceleration
$M_{\roman E}/{R_{\roman E}\!\!}^2$, Earth's surface gravitational
acceleration of about 9.8 m s$^{-2}$.  A cargo weighing many tons would
constitute an ether sink inside the spaceship, but one whose gravitational
escape velocity would be negligible in comparison to Earth's.  It would
increase the ether velocity and acceleration above the vessel by relatively
insignificant amounts, and would decrease them by similar amounts below the
vessel.  The change in burden on the drive would be slight.  On the other
hand, to progress from supporting the vessel against the Earth's gravitational
pull to zooming it along through space at the speed of light could require a
considerable increase in the drive's efficiency.  Its tiny topological
tunnels would be required to pass ether at $3.0 \times 10^8$ m s$^{-1}$,
instead of a paltry $1.1 \times 10^4$ m s$^{-1}$.  Maintaining the velocity
through the tunnels at the speed of light, once attained, could be no more
taxing than allowing a flowing river to keep on flowing.  Attaining that
velocity in a reasonable time would present the difficulty.  To go from rest
to traveling at lightspeed in an hour, a day, or a year would, according to
the time-honored formula $v = a t$, demand that the ether be constantly
accelerated through the tunnels at about $8.3 \times 10^4$ m s$^{-2}$,
$3.5 \times 10^3$ m s$^{-2}$, or 9.5 m s$^{-2}$, respectively.  The ether
acceleration just sufficient for levitation at Earth's surface thus would
produce lightspeed only after about one year.  This acceleration, if
sustained for 50 light-years and reversed for another 50, would cause a
one-way trip of 100 light-years to last 20 (= $2 \sqrt{2 \cdot 50}$) years.
If lightspeed could be attained in a day, the time required would be
$20/\sqrt{365}$ ($\approx 1.05$) years; if in an hour, $20/\sqrt{8760}$ years
($\approx 78$ days).

The maximum acceleration of the ether through a static drainhole occurs
where the throat is most constricted, that is, at $\rho = 2m$, where $r$
has its minimum value.  The magnitude of the acceleration there is
$|m|/r^2 (2m)$, a number of the order of $|m|/n^2$ whatever the relative
sizes of $m$ and $n$.  To put a face on this number, suppose
$m = m_{\roman {neutron}} \approx 1.2 \times 10^{-54}$ m (the rest mass --- and
{\it presumably} a close approximation to the active gravitational mass ---
of the neutron in units in which $G = c = 1$, which have been assumed here) and
$n = n_{\roman {Pla}} \approx 1.6 \times 10^{-35}$ m (the Planck distance).
From (12) one finds that for light rays
$(d\rho/dt - v_{\roman e}\!{}^\rho (\rho))^2 +
r^2 (\rho) (d\vartheta/dt)^2 +
r^2 (\rho) (\sin \vartheta)^2 (d\varphi/dt)^2 = 1$, so that the speed
of light with respect to the ether, thus with respect to an observer in
radial free fall through the drainhole, is indeed 1.  This makes
$1\; \text{m} \approx 3.3 \times 10^{-9}$ s, so that
$n_{\roman {Pla}} \approx 5.3 \times 10^{-44}$ s, and therefore
$m/n^2 \approx 4.3 \times 10^{32}$ m s$^{-2}$.  If, on the other hand,
$n = n_{\roman {electron}} \approx 2.8 \times 10^{-15}$ m (the classical radius
of the electron), then $m/n^2 \approx 1.3 \times 10^{-8}$ m s$^{-2}$.  Thus
tiny topological tunnels of radius $n_{\roman {Pla}}$ would likely produce
a far stronger drive than tunnels of radius $n_{\roman {electron}}$,
{\it provided} they could be created and distributed in quantities
sufficient to divert all of the ether flowing into the spaceship's hull.

At this point we are off the end of the good highway built on firm
mathematics, and at risk of wandering lost in the desert.  We had best
retreat to the pavement and see what can be done to extend it beyond the
horizon.  Perhaps we shall be able only to survey a mathematical route on
which we can never pour the concrete of physical existence.  Even so, a
warp drive would be such a  marvelous thing to possess that one cannot help
longing for its creation.  Should we not set our minds to the task?  And if
we are not alone in the Universe, is it not likely that others already have
set theirs, perhaps to good effect?

\enddocument